\documentclass{article}  
\usepackage{amsmath,amsfonts,amsthm,fullpage}

\newsavebox{\fmbox}
\newsavebox{\algobox}
\newenvironment{algo}[1]
{\begin{center}\begin{lrbox}{\algobox}\begin{minipage}{#1}}
{\end{minipage}\end{lrbox}\fbox{\usebox{\algobox}}\end{center}}

\newtheorem{theorem}{Theorem}
\newtheorem{definition}{Definition}
\newtheorem{corollary}{Corollary}
\newtheorem{lemma}{Lemma}
\newtheorem{fact}{Fact}

\newcommand{\ket}[1]{{|{#1}\rangle}}
\newcommand{\size}[1]{\lvert #1 \rvert}
\renewcommand{\epsilon}{\varepsilon}

\begin{document}

\title{Quantum Algorithms for the Triangle Problem\thanks{A 
preliminary version of this paper appeared in
\emph{Proceedings of 16th ACM-SIAM Symposium on Discrete Algorithms}, 
pp. 1109--1117, 2005.}}
\author{Fr\'ed\'eric Magniez\thanks{
CNRS--LRI, UMR 8623 Universit\'e Paris--Sud,
91405 Orsay, France,
emails:  \{{\tt magniez}, {\tt santha}\}{\tt @lri.fr};
partially supported by
the EU 5th and 6th framework programs
RESQ IST-2001-37559, RAND-APX IST-1999-14036
and QAP IST,
and by ACI Cryptologie CR/02 02 0040 
and ACI S\'ecurit\'e Informatique 03 511
grants of the French Research Ministry.
}
\and
Miklos Santha${}^{*}$
\and
Mario Szegedy\thanks{
Rutgers University,
email: {\tt szegedy@cs.rutgers.edu}; 
supported by NSF grant 0105692
and the EU 5th framework program
RESQ IST-2001-37559. The research was done while the author was visiting LRI.
}
}
\date{}

\maketitle

\begin{abstract}
We present two new quantum algorithms that either find a triangle (a copy of $K_{3}$)
in an undirected graph $G$ on $n$ nodes, or reject if $G$ is triangle free.
The first algorithm uses combinatorial ideas with Grover Search
and makes $\tilde{O}(n^{10/7})$ queries.
The second algorithm uses $\tilde{O}(n^{13/10})$ queries, and it is based on
a design concept of Ambainis~\cite{amb04} that 
incorporates the benefits of quantum walks into Grover search~\cite{gro96}.
The first algorithm uses only $O(\log n)$ qubits in its
quantum subroutines, whereas the second one uses $O(n)$ qubits.
The Triangle Problem was first treated in~\cite{bdhhmsw01}, where an
algorithm with $O(n+\sqrt{nm})$ query complexity was presented,
where $m$ is the number of edges of $G$.
\end{abstract}

\section{Introduction}
Quantum computing is an extremely active research area
(for introductions see e.g.~\cite{nc00,ksv02})
where a growing trend is the study of quantum query complexity.
The quantum query model was implicitly introduced by Deutsch, Jozsa,
Simon and Grover~\cite{deu85,dj92,sim97,gro96}, and explicitly by
Beals, Buhrman, Cleve, Mosca and de Wolf~\cite{bbcmw01}.
In this model, like in its classical counterpart, we pay 
for accessing the oracle (the black box), but unlike in the classical case,
the machine can use the power of quantum parallelism 
to make queries in superpositions.
While no significant lower bounds are known in quantum time complexity,
the black box constraint sometimes enables us to prove such bounds
in the query model.

For promise problems
quantum query complexity indeed can be exponentially smaller
than the randomized one, a prominent example for that is the
Hidden Subgroup Problem~\cite{sim97,ehk99}.
On the other hand,
Beals, Buhrman, Cleve, Mosca and de Wolf~\cite{bbcmw01} showed that
for total functions the deterministic and the quantum
query complexities are polynomially
related.
In this context, a large axis of research pioneered by Grover~\cite{gro96}
was developed around search problems in unstructured, structured, or
partially structured databases.

The classical query complexity of graph properties 
has made its fame through the notoriously hard evasiveness
conjecture of Aanderaa and Rosenberg~\cite{ros73}
which states that every non-trivial and monotone boolean function on graphs 
whose value remains invariant 
under the permutation of the nodes has
deterministic query complexity exactly ${ n\choose 2}$, where $n$
is the number of nodes of the input graph.
Though this conjecture is still open, an $\Omega (n^2)$ lower bound
has been established by Rivest and Vuillemin~\cite{rv76}.
In randomized bounded error complexity
the general lower bounds are far from the 
conjectured $\Omega(n^{2})$. 
The first non-linear lower bound was shown by Yao~\cite{yao87}. 
For a long time Peter Hajnal's $\Omega(n^{4/3})$ bound \cite{haj91} 
was the best, until it was slightly
improved in \cite{ck01} to 
$\Omega(n^{4/3} \log^{1/3}n)$.
The question of the
quantum query complexity of graph properties was first 
raised in \cite{bcwz99} where it is shown that 
in the exact case
an $\Omega(n^{2})$ lower bound still holds.
In the bounded error quantum query model,
the $\Omega(n^2)$ lower bound does not hold anymore in general.
An  $\Omega(n^{2/3} \log^{1/6}n)$ lower bound, first observed by Yao~\cite{yao03},
can be obtained combining Ambainis' technique~\cite{amb02} with the above randomized lower bound.

We address the Triangle Problem in this setting.
In a graph $G$,
a complete subgraph on three vertices is called a {\em triangle}.
In this write-up we study the following oracle problem:
\begin{quote}
\textsc{Triangle}\\
\textit{Oracle Input:} The adjacency matrix $f$ of a graph $G$
on $n$ nodes. \\
\textit{Output:} a triangle if there is any,
otherwise reject.
\end{quote}
\textsc{Triangle} has been studied in various contexts,
partly because of its relation to matrix multiplication~\cite{ayz97}.
Its quantum query complexity was first raised in~\cite{bdhhmsw01},
where the authors show that in the case of sparse graphs 
the trivial (that is, using Grover Search) $O(n^{3/2})$ upper bound  can be improved.
Their method breaks down when
the graph has $\Theta(n^{2})$ edges.

The quantum query complexity of \textsc{Triangle}
as well as of many of its kins with small 
one-sided certificate size are notoriously hard to analyze, because
one of the main lower bounding methods breaks down near 
the square root of the instance size~\cite{sze03,lm04,zha04,ss05}:
{\em
If the $1$-certificate size of a boolean function 
on $N$ boolean variables is $K$,
then even the most general variants~\cite[Theorem~4]{bss03}\cite{amb03:vs}\cite{lm04} of the 
Ambainis' quantum adversary technique~\cite{amb02} can prove only a lower bound of $\Omega(\sqrt{NK})$.}
Indeed only the $\Omega(n)$  lower bound is known for \textsc{Triangle}, 
which, because of the remark above,
cannot be improved using any quantum adversary technique ($N=n^2$ and $K=3$).
Problems with small certificate complexity include various 
collision type problems such as the  2-1 Collision Problem
and the Element Distinctness Problem.
The first polynomial lower bound for the 2-1 Collision Problem  was shown by Aaronson
and Shi~\cite{as04}
using the polynomial method of Beals, Buhrman, Cleve, Mosca and de Wolf \cite{bbcmw01}.
For the Element Distinctness Problem, 
a randomized reduction from the 2-1 Collision Problem gives
$\Omega(n^{2/3})$.

In this paper we present two different approaches that give rise 
to new upper bounds.
First, using combinatorial ideas, 
we design an algorithm for \textsc{Triangle} (\textbf{Theorem~\ref{mario:algo}})
whose quantum query complexity is $\tilde{O}(n^{10/7})$.  
Surprisingly, its quantum parts only consist in Grover Search subroutines.
Indeed, Grover Search coupled with the Szemer\'edi Lemma~\cite{sze76}
already gives a $o(n^{3/2})$ bound. We exploit this fact using a simpler
observation that leads to the $\tilde{O}(n^{10/7})$ bound.
Moreover our algorithm uses only small quantum memory, namely $O(\log n)$ qubits
(and $O(n^2)$ classical bits).
Then, we generalize 
the new elegant method used by Ambainis~\cite{amb04} for solving
the Element Distinctness Problem in $O(n^{2/3})$, to solve
a general Collision Problem by a dynamic quantum query algorithm
(\textbf{Theorem~\ref{ambainis1}}). 
The solution of the general Collision Problem will be used in our 
second algorithm for \textsc{Triangle}.
As an intermediate step, we introduce the Graph Collision Problem, which
is a variant of the Collision Problem, and solve it in $\tilde{O}(n^{2/3})$ query complexity
(\textbf{Theorem~\ref{graph-collision}}).
Whereas a reduction of  \textsc{Triangle} to the Element Distinctness Problem does not
give a better algorithm than $O(n^{3/2})$, 
using a recursion of our dynamic version of Ambainis' method we prove the $\tilde{O}(n^{13/10})$
query complexity for \textsc{Triangle} (\textbf{Theorem~\ref{triangle}}).
We end by generalizing this result 
for finding the copy of any given graph (\textbf{Theorem~\ref{h-copy}})
and then for every graph property with small
$1$-certificates (\textbf{Corollary~\ref{certificate}}).

\section{Preliminaries}
\subsection{Query Model}
In the query model of computation each query adds one to
the complexity of an algorithm, but
all other computations are free.
The state of the computation is represented by three
registers, the query register $x$, the answer register $a$, and the
work register $z$. The computation takes place in the vector space spanned by all
basis states $\ket{x,a,z}$.
In the {\em quantum query model} the state of the computation is a complex
combination of all basis states which has unit length in the norm $l_2$.

The query operation $O_f$ maps the
basis state
$\ket{x,a,z}$
into the state $\ket{x,a\oplus f(x),z}$
(where $\oplus$ is bitwise XOR).
Non-query operations are independent of $f$.
A {\em $k$-query algorithm} is a sequence of $(k+1)$ operations
$(U_0, U_1, \ldots , U_k)$ where $U_i$ is unitary.
Initially the state of the computation is set to some
fixed value $\ket{0,0,0}$, and then the sequence of operations
$U_0, O_f, U_1, O_f, \ldots, U_{k-1}, O_f, U_k$ is applied.

\subsection{Notations}
We denote the set $\{1,2,\ldots,n\}$ by $[n]$.
A simple undirected graph is a set of edges $G\subseteq
\{(a,b)\mid \; a,b\in [n];\; a\neq b\}$ with the understanding that
$(a,b) \stackrel{\text{def}}{=} (b,a)$. 
Let $t(G)$ denote the number of triangles in $G$.
The complete graph on a set $\nu\subseteq [n]$ is denoted by $\nu^{2}$.
The neighborhood of a $v\in [n]$
in $G$ is denoted by $\nu_{G}(v)$, and it is defined by
$\nu_{G}(v) = \{b\mid \; (v,b)\in G\}$.
We denote $|\nu_{G}(v)|$ by $\deg_{G} v$.
For sets $A,B\subseteq [n]$ let
$G(A,B) =  \{(a,b)\mid \; a\in A;\;b\in B;\;(a,b)\in G\}.$

The following function will play a major role in our proof.
We denote the number of paths of length two from $a\in [n]$
to $b\in [n]$ in $G$ with $t(G,a,b)$:
$t(G,a,b) = |\{x\mid \; (a,x)\in G;\; (b,x)\in G \}|$.
For a graph $G\subseteq [n]^{2}$ and an integer $k\geq 0$, we define
$G^{\langle k \rangle} = \{ (a,b)\in [n]^{2}\mid \; t(G,a,b) \le k \}.$

\subsection{Quantum Subroutines}
We  will use a safe version of Grover Search~\cite{gro96}, namely
\textbf{Safe Grover Search$(t)$},
based on a $t$ iterations of Grover Search,
and followed by a checking process for markedness of of output instances.
\begin{fact}
Let $c>0$.
\textbf{Safe Grover Search$(\Theta(c\log N))$} on a database of $N$ items 
has quantum query complexity $O(c\sqrt{N} \log N)$ and 
it always rejects if there is no marked item, otherwise it
finds a marked item with probability at least $1-{1\over N^{c}}$.
\end{fact}

For quantum walks on graphs we usually define two operators:
{\em coin flip} and {\em shift}. The state of the walk is held in a 
pair of registers, the {\em node} and the {\em coin}.
The coin flip operator acts only on the coin register and it is the identity on the node register.
The shift operation only changes the node register, 
but it is controlled by the content of the coin register
(see~\cite{wat01,aakv01,abnvw01}).
Often the coin flip is actually the Diffusion operator.
\begin{definition}[Diffusion over $T$]
Let $T$ be a finite set. 
The {\em diffusion} operator over $T$ is
the unitary operator on the Hilbert space 
${\bf C}^{T}$ that acts on a basis element $|x\rangle$, $x\in T$ as:
$\ket{x}\mapsto -\ket{x}+\tfrac{2}{\size{T}}\sum_{y\in T}\ket{y}$.
\end{definition}

In~\cite{amb04} a new walk is described
that plays a central role in our result.
Let $S$ be a finite set of size $n$.
The node register holds 
a subset $A$ of $S$ of size either $r$ or $r+1$
for some fixed $0 <  r<n$, and the coin register 
holds an element $x\in S$. Thus the basis states are
of the form $\ket{A}\ket{x}$, where we also require that
if $|A|=r$ then $x\not\in A$, and if $|A|=r+1$ then
$x\in A$. We also call the node register the {\em set register}.

\begin{algo}{7cm}
\textbf{Quantum Walk}\smallskip
\begin{enumerate}\setlength{\itemsep}{0pt} 
\item Diffuse the coin register over $S-A$
\item Add $x$ to $A$ 
\item Diffuse the coin register over $A$
\item Remove $x$ from $A$
\end{enumerate}
\end{algo}

Ambainis~\cite{amb04} showed that, inside some specific stable subspaces,
$\Theta(\sqrt{r})$ iterations of \textbf{Quantum Walk} can play the role of 
the diffusion over $\{(A,x) : A\subset S, \size{A}=r, x\not\in S\}$.
This nice result leads to a more efficient Grover search for some problems
like the Element Distinctness Problem~\cite{amb04}.
We will describe this in a general setting in Section~\ref{section:ambainis}.

\section{Combinatorial Approach}
\subsection{Preparation}
The algorithm presented here is based on three combinatorial observations.
Throughout this section we
do not try to optimize $\log n$ factors and we will hide time in the
$\tilde{O}$ notation.
The first observation
is based on the Amplitude Amplification technique
of Brassard, H{\o}yer, Mosca, and Tapp~\cite{bhmt02}
\begin{lemma}\label{many}
For any known graph $E\subseteq [n]^2$, a triangle with at least one edge 
in $E$ can be detected with $\tilde{O}(\sqrt{E}+\sqrt{n|G\cap E|})$ queries
and probability $1-\tfrac{1}{n}$.
\end{lemma}

Perhaps the most crucial observation to the algorithm is the following
simple one.
\begin{lemma}\label{trivi}
For every $v\in [n]$, using
$\tilde{O}(n)$ queries, we either find a triangle in $G$ or verify that 
$G\subseteq [n]^{2}\setminus \nu_{G}(v)^{2}$ with probability $1 - 
{1\over n^3}$.
\end{lemma}
\begin{proof}
We query all edges incident to $v$ classically
using $n-1$ queries. This determines $\nu_{G}(v)$.
With Safe Grover Search we find an edge of $G$ in
$\nu_{G}(v)^{2}$, if there is any.
\end{proof}

This lemma with the observation that hard instances 
have to be dense, already enable us to show that the quantum 
query complexity of \textsc{Triangle} is $o(n^{3/2})$, using 
the Szemer\'edi Lemma~\cite{sze76}. However another 
fairly simple observation can help us to decrease the exponent.
\begin{lemma}\label{almosttrivi}
Let $0<\epsilon<1$, $k = \lceil 4n^{\epsilon}\log n\rceil$, and let
$v_{1},v_{2},\ldots,v_{k}$
randomly chosen from $[n]$ (with no repetitions).
Let
$G' = [n]^{2}\setminus \cup_{i=1}^{k} \nu_{G}(v)^{2}$.
Then
$\Pr_{v_{1},v_{2},\ldots,v_{k}}\left( G'\subseteq G^{\langle 
n^{1-\epsilon}\rangle }   \right) > 1- {1\over n}.$
\end{lemma}

Let us first remind the reader 
about the following lemma that is useful in many applications.
\begin{lemma}\label{useful}
Let $X$ be a fixed subset of $[n]$ of size $pn$ and $Y$ be a random subset
of $[n]$ of size $qn$, where $p+q<1$. Then the probability that 
$X\cap Y$ is empty is $(1- pq)^{n(1\pm O(p^{3}+q^{3}+1/n))}$.
\end{lemma}

\begin{proof}
The probability we are looking for 
is estimated using the Stirling formula as
\begin{eqnarray*}
{{n(1-p)\choose nq}\over {n\choose nq} } &= &
{[n(1-p)]! [nq]! [n(1-q)]! \over [nq]! [n(1-p-q)]!n!}\\
&=&\sqrt{ \tfrac{(1-p)(1-q)}{ 1-p-q}}
\left[
\tfrac{(1-p)^{1-p} (1-q)^{1-q}}{(1-p-q)^{1-p-q}}\right]^{n}{\scriptstyle (1\pm o(1))}\\
& = & (1- pq)^{n(1 \pm O(p^{3}+q^{3}+1/n))}.
\end{eqnarray*} 
\end{proof}

\begin{proof}[Proof of Lemma~\ref{almosttrivi}]
Consider now a fixed edge $(a,b)$ such that
$t(G,a,b) \ge  n^{1-\epsilon}$. The probability that 
$(a,b)\in G'$ is the same as the probability that the set 
$X = \{x\in [n] : (x,a)\in G \text{ and }(x,b)\in G\}$
is disjoint from the 
random set $\{v_{1},v_{2},\ldots,v_{k}\}$. Notice that 
$|X|= t(G,a,b)$. By Lemma \ref{useful} we can estimate now this probability
as, for sufficiently large $n$,
$$
\left(1 - {4n^{\epsilon}\log n\over n}\times
{n^{1 - \epsilon}\over n}\right)^{
n(1+o(1))}
= \left( 1 - {4\log n\over n}\right)^{n(1+o(1))}
< e^{-3\log n} = n^{-3}.$$
Then the lemma follows from the union bound, since the 
number of possible edges $(a,b)$ is at most $n^{2}$.
\end{proof}

\subsection{Algorithm and Analysis}
We now describe our algorithm where every searches are done
using \textbf{Safe Grover Search}.
We delay details of Step~\ref{classification} for a while.

\begin{algo}{12cm}
\textbf{Combinatorial Algorithm$(\epsilon,\delta,\epsilon')$}\smallskip
\begin{enumerate}\setlength{\itemsep}{0pt} 
\item Let $k=\lceil 4 n^{\epsilon}\log n\rceil$
\item Randomly choose $v_1,\ldots,v_k$ from $[n]$ (with no repetition)
\item Compute every $\nu_G(v_i)$
\item If  $G\cap \nu_G(v_i)^2\neq\emptyset$, for some $i$, then output the triangle
induced by $v_i$
\item Let $G'=[n]^2\setminus \cup_i (\nu_{G}(v_{i})^{2})$
\item Classify the edges of $G'$ into $T$ and $E$ such that\label{classification}\\
\quad -- $T$ contains only $O(n^{3-\epsilon'})$ triangles\\
\quad -- $E\cap G$ has size $O(n^{2-\delta}+n^{2-\epsilon+\delta+\epsilon'})$
\item Search for a triangle in $G$ among all triangles inside $T$\label{search}
\item Search for a triangle of $G$ intersecting with $E$
\item Output a triangle if it is found, otherwise reject
\end{enumerate}
\end{algo}

\begin{theorem}\label{mario:algo}
\textbf{Combinatorial Algorithm$(\epsilon,\delta,\epsilon')$}
rejects with
probability one if there is no triangle in $G$, otherwise returns a
triangle of $G$ with probability $1-O(\frac{1}{n})$.
Moreover it has query complexity
$\tilde{O}\left(
n^{1+\epsilon} + 
n^{1+ \delta + \epsilon'}  + 
\sqrt{n^{3-\epsilon'}} +
\sqrt{n^{3-\min(\delta,\epsilon-\delta-\epsilon')}}\right)$.
\end{theorem}
With $\epsilon={3\over 7}$, $\epsilon'=\delta={1\over 7}$
this gives $\tilde{O}(n^{1+{3\over 7}})$ for the total number of queries.

We require every probabilistic steps to be correctly performed
with probability $1-O(\tfrac{1}{n^3})$. So that the overall
probability of a correct execution is $1-O(\tfrac{1}{n})$, using
the union bound and since the number of such steps is at most $O(n^2)$.
Thus we will always assume that an execution is correct.
Since an incorrect execution might increase the query complexity
of the algorithm, we also assume there is a counter so that the
algorithm rejects and stops when a threshold is exceeded.
This threshold is defined as the maximum of
query complexities over every correct executions.

The main step of \textbf{Combinatorial Algorithm} is Step~\ref{classification} that
we implement in the following way.
\vfill

\begin{algo}{13cm}
\textbf{Classification$(G',\delta,\epsilon')$}\smallskip
\begin{enumerate}\setlength{\itemsep}{0pt} 
\item Set $T=\emptyset$, $E=\emptyset$
\item While $G'\neq\emptyset$ do\vspace*{-\parsep}
\begin{enumerate}\setlength{\itemsep}{0pt}
\item While there is an edge $(v,w)\in G'$ s.t. $t(G',v,w) <  n^{1-\epsilon'}$\\
\phantom{--} Add $(v,w)$ to $T$, and delete it from $G'$\label{stept}
\item Pick a vertex $v$ of $G'$ with non-zero degree and 
decide\label{strategy}\\
\hspace*{-0.5cm}{\em 1. low degree hypothesis}: $|\nu_{G}(v)| \le 
10  \times n^{1-\delta}$\\
\hspace*{-0.5cm}{\em 2. high degree hypothesis}: $|\nu_{G}(v)| \ge 
\tfrac{1}{10} \times n^{1-\delta}$
\item If Hypothesis 1, add all \label{low}
edges $(v,w)$ of $G'$ to $E$, and delete them from $G'$
\item If Hypothesis 2, then\vspace*{-\parsep} \label{high}
\begin{enumerate}\setlength{\itemsep}{0pt}
\item Compute $\nu_G(v)$
\item If $G\cap\nu_{G}(v)^2\neq\emptyset$, output the triangle induced by $v$ and stop
\item Add all edges in $G'(\nu_{G}(v),\nu_{G'}(v))$ to $E$, and
delete them from $G'$
\end{enumerate}\vspace*{-\parsep}
\end{enumerate}\vspace*{-\parsep}
\end{enumerate}
\end{algo}
In Step~\ref{strategy}, we use an obvious sampling strategy:
\begin{quote}
Set a counter $C$ to 0.
Query $\lceil n^{\delta}\rceil$ random edge candidates 
from $v\times [n]$.
If there is an edge of $G$ among them, 
add one to $C$. Repeat this process $K= c_{0}\log n$ times,
where $c_{0}$ is a sufficiently large constant.
Accept the low degree hypothesis if by the end $C<K/2$,
otherwise accept the large degree hypothesis. 
\end{quote}
Observe than one could use here a quantum procedure based on Grover Search.
Since the cost of this step is negligible from others, this would not give any better bound.
\begin{fact}\label{firstfact} When $c_0$ is large enough in Step~\ref{strategy}:
\begin{enumerate}\setlength{\itemsep}{0pt} 
\item The probability that $\deg_{G}(v) > 10\times n^{1-\delta}$
and the low degree hypothesis is accepted is  $O(\tfrac{1}{n^{3}})$.
\item The probability that $\deg_{G}(v) < \tfrac{1}{10}\times n^{1-\delta}$
and the high degree hypothesis is accepted is $O(\tfrac{1}{n^{3}})$.
\end{enumerate}
\end{fact}
\begin{proof}
Indeed, using Lemma \ref{useful}, considering a single 
round of sampling the probability that our sample set 
does not contain an edge from $G$ even though
$\deg_{G}(v) > 10\times n^{1-\delta}$ is, for sufficiently large $n$,
$$\left(1 - {10n^{1-\delta}\over n}\times
{n^{\delta}\over n}\right)^{
n(1+o(1))} =\left(1 - {10\over n}\right)^{
n(1+o(1))} < 0.1 .$$
Similarly, the probability that our sample set 
contains an edge from $G$ even though
$\deg_{G}(v) < \tfrac{1}{10}\times n^{1-\delta}$ is 
$$1- \left(1 - {n^{1-\delta}\over 10n}\times
{n^{\delta}\over n}\right)^{
n(1+o(1))} = 1- \left(1 - {1\over 10n}\right)^{
n(1+o(1))} < 0.2 .$$

Now for $K= c_{0}\log n$ rounds, where $c_{0}$ is large enough,
the Chernoff bound gives the claim.
\end{proof}

\begin{lemma}\label{lemmaclassification}
If $G\subseteq G'\subseteq  G^{\langle 
n^{1-\epsilon}\rangle }$, then \textbf{Classification$(G',\epsilon',\delta)$}
output the desired partition $(T,E)$ of $G$ with probability
$1-O(\tfrac{1}{n})$ and has query complexity
$\tilde{O}(n^{1+\delta+\epsilon'})$.
\end{lemma}

\begin{proof}[Proof of Theorem~\ref{mario:algo}]
Clearly, if there is no triangle in the graph,
the algorithm rejects since the algorithm
outputs a triplet only after checking that it is a triangle in $G$.
Therefore the correctness proof requires only to calculate 
the probability with which the algorithm outputs a triangle
if there is any, and the query complexity of the algorithm.

Assume that the execution is without any error.
Using union bound, we can indeed upper bounded the probability of
incorrect execution by $O(\tfrac{1}{n})$.

By Lemma~\ref{trivi}, we already know that the construction of $G'$
requires $\tilde{O}(n^{\epsilon}\times n)$ queries. 
Moreover either $G\subseteq G'$ or a triangle is found, with probability
$1-O(\tfrac{1}{n})$.
{From} Lemma~\ref{almosttrivi}, we also know that
$G' \subseteq  G^{\langle n^{1-\epsilon}\rangle }$ with
probability $1-O({1\over n})$.

Assume that $G'$ lends all its edges 
to $T$ and $E$, that is no triangle is found at the end of \textbf{Classification}.
Since $G\subseteq G'$,
every triangle in $G$ either has to be contained totally in $T$
or it has to have a non-empty intersection with $E$.
Using Lemma~\ref{lemmaclassification}, we know that the partition $(T,E)$
is correct with probability $1-O(\tfrac{1}{n})$. Assume this is the case.
$T$ is a graph that is known to us, and so we can
find out if one of these triangles belong to $G$ 
with $\tilde{O}(\sqrt{n^{3-\epsilon'}})$ queries, using \textbf{Safe Grover Search}.
By Lemma~\ref{many}, the complexity of finding a triangle in $G$
that contains an edge from $E$  is 
$\tilde{O}\left(n + \sqrt{n^{3-\min(\delta,\epsilon
-\delta-\epsilon')}}\right)$.

{From} the analysis we conclude that the total 
number of queries is upper bounded by:
$$\tilde{O}\left( n^{1+\epsilon} + 
n^{1+\epsilon} + 
(n^{1+ \delta + \epsilon'} + n^{1+\delta}) + 
\sqrt{n^{3-\epsilon'}}
 +\sqrt{n^{3-\min(\delta,\epsilon-\delta-\epsilon')}}\right).$$
\end{proof}

In the rest of the section we prove Lemma~\ref{lemmaclassification}
using a sequence of facts. 
Then the proof derives directly
noting that Step~\ref{high} has query complexity $\tilde{O}(n)$.

\begin{fact}\label{facthigh}
During a correct execution, there is at most
${O}(n^{\delta+\epsilon'})$ iterations of Step~\ref{high}.
\end{fact}
\begin{proof}
We will estimate the number of executions of Step~\ref{high}
by lower bounding $|G'(A,A')|$, where
$A=\nu_G(v)$ and $A'=\nu_{G'}(v)$. For each $x\in A$
we have $t(G',v,x) \ge n^{1-\epsilon'}$, otherwise
in Step~\ref{stept} we would have classified $(v,x)$ into $T$.
A triangle $(v,x,y)$ contributing to 
$t(G',v,x)$ contributes with the edge $(x,y)$ to $G'(A,A')$.
Two different triangles $(v,x,y)$ and $(v,x',y')$
can give the same edge in $G'(A,A')$ only if $x=y'$ and $y=x'$.
Thus:
\begin{equation}\label{trieq}
|G'(A,A')| \ge {1\over 2} \sum_{x\in\nu_{G}(v)} t(G',v,x)
\ge   |A| n^{1-\epsilon'}/2.
\end{equation}

Since we executed Step~\ref{high} only under the large degree
hypothesis on $v$, if the hypothesis is correct,
the right hand side of Equation \ref{trieq} is at least
$\tfrac{1}{10}\times n^{1-\delta}\times
n^{1-\epsilon'}/2 = \Omega(n^{2-\delta - \epsilon'})$.
Since $G'$ has at most ${n\choose 2}$ edges, 
it can execute Step~\ref{high} at most $O(n^{\delta + \epsilon'})$ times.
\end{proof}

\begin{fact}\label{factlow}
During a correct execution, there is at most
$O(n)$ iterations of Step~\ref{low}.
\end{fact}
\begin{proof}
We claim that 
each vertex is processed in Step~\ref{low} at most once.
Indeed, if a vertex $v$ gets into Step~\ref{low}, its incident edges
are all removed, and its degree in $G'$ becomes 0
making it ineligible for being processed in Step~\ref{low} again.
\end{proof}

Now we state that $T$ contains $O(n^{3-\epsilon'})$ triangles
using this quite general fact.
\begin{fact}\label{trianglefact}
Let $H$ be a graph on $[n]$. Assume that a graph $T$ is built
by a process that starts with an empty set, and
at every step either discards some edges from
$H$ or adds an edge $(a,b)$ of $H$ to $T$ for which
$t(H,a,b)\le\tau$ holds. For the $T$ created by the end of the process
we have $t(T)\le { n \choose 2}\tau$.
\end{fact}
\begin{proof}
Let us denote by $T[i]$ the edge of $T$
that $T$ acquired when it was incremented for 
the $i^{\rm th}$ time, and let us use the notation $H^{i}$ 
for the current version of $H$ before the very 
moment when $T[i] = (a_{i},b_{i})$
was copied into $T$. Since 
$\{T[i],T[i+1],\ldots \} \stackrel{\text{def}}{=} T^{i} \subseteq H^{i}$,
we have 
$t(T^{i},a_{i},b_{i}) 
\le t(H^{i},a_{i},b_{i}) \le \tau.$
Now the fact follows from
$t(T) = \sum_{i} t(T^{i},a_{i},b_{i}) \le 
{ n \choose 2}\tau,$
since $i$ can go up to at most ${ n \choose 2}$.
\end{proof}

\begin{fact}
During a correct execution,
$E\cap G$ has size $O(n^{2-\delta}+n^{2-\epsilon+\delta+\epsilon'})$.
\end{fact}
\begin{proof}
In order to estimate $E\cap G$ observe that we
added edges to $E$ only in Steps~\ref{low} and~\ref{high}.
In each execution of Step~\ref{low}, we added at most 
$10 n^{1-\delta}$ edges to $E$, and we had $O(n)$
such executions (Fact~\ref{factlow}) that give a total of 
$O(n^{2-\delta})$ edges. The number of executions of Step~\ref{high}
is $O(n^{\delta + \epsilon'})$ (Fact~\ref{facthigh}). Our task is now to bound
the number of edges of $G$ each such execution
adds to $E$.

We estimate $|G\cap G'(A,A')|$ from the $A'$ side, where
$A=\nu_G(v)$ and $A'=\nu_{G'}(v)$.
This is the only place where we use the fact that 
$G' \subseteq  G^{\langle 
n^{1-\epsilon}\rangle }$: For every $x\in A'$ we have 
$t(G,v,x)\le n^{1-\epsilon}$.
On the other hand, when $y\in A$ and $x\in A'$, 
every edge $(y,x)\in G'$
creates a $(v,x)$-based triangle.
Thus 
$$|G\cap G'(A,A')| \le |A'| n^{1-\epsilon} \le n^{2-\epsilon}.$$

Therefore the total number of edges of $G$ Step~\ref{high}
contributes to $E$ is $n^{2-\epsilon+\delta+\epsilon'}$.
In conclusion,
$$|G\cap E|\le O(n^{2-\delta}+n^{2-\epsilon+\delta+\epsilon'}).$$
\end{proof}

\section{Quantum Walk Approach}

\subsection{Dynamic Quantum Query Algorithms}\label{section:ambainis}

The algorithm of Ambainis in~\cite{amb04} 
is somewhat similar to the brand of classical
algorithms, where a database is used (like in heapsort)
to quickly retrieve the value of those items needed for the run of the 
algorithm. Of course, this whole paradigm is placed 
into the context of query algorithms. We shall define 
a class of problems that can be tackled very well with 
the new type of algorithm. Let $S$ be a finite set of size $n$
and let $0< k < n$.

\begin{quote}
\textsc{$k$-Collision}\\
\textit{Oracle Input:} A function $f$ which defines a relation $\mathcal{C}\subseteq S^k$. \\
\textit{Output:} 
A $k$-tuple $(a_1,\ldots,a_k)\in\mathcal{C}$ if it is non-empty,
otherwise reject.
\end{quote}

By carefully choosing the relation $\mathcal{C}$, \textsc{$k$-Collision}
can be a useful building block in the design of different algorithms. 
For example if $f$ is the adjacency
matrix of a graph $G$, and the relation $\mathcal{C}$ is
defined as
`being an edge of a triangle of $G$' then the output of \textsc{Collision}
yields a solution for  \textsc{Triangle} with $O(\sqrt{n})$ additional queries (Grover search for the
third vertex).
\begin{quote}
\textsc{Unique $k$-Collision}:
The same as \textsc{$k$-Collision} with the promise that 
$|\mathcal{C}|=1$ or $|\mathcal{C}|=0$.
\end{quote}

The type of algorithms we study will use a database $D$
associating some data
$D(A)$ to every set $A\subseteq S$.
{From} $D(A)$ we would like to determine
if $A^k\cap \mathcal{C}\neq\emptyset$. We expedite this using 
a quantum query procedure $\Phi$ with the property that
$\Phi(D(A))$ rejects if $A^k\cap \mathcal{C}=\emptyset$
and, otherwise, both accepts and outputs an element of $A^k\cap \mathcal{C}$, that
is a {\em collision}.
When operating with
$D$ three types of costs incur, all measured in the number of queries to the oracle $f$.

\begin{description}\setlength{\itemsep}{0pt}
\item[Setup cost $s(r)$:] The cost to set up $D(A)$
for a set of size $r$.
\item[Update cost $u(r)$:]  The cost to update $D$ for a set of size $r$, i.e. moving from $D(A)$ 
to $D(A')$, where $A'$ results from $A$ by adding an element,
or moving from $D(A'')$ to $D(A)$ where $A$ results from $A''$ by deleting an element.
\item[Checking cost $c(r)$:] The query complexity of
$\Phi(D(A))$ for a set of size $r$.
\end{description}

Next we describe the algorithm of Ambainis~\cite{amb04} in general terms.
The algorithm has 3 registers $\ket{A}\ket{D(A)}\ket{x}$.
The first one is called the {\em set register}, the second one the {\em data register},
and the last one the {\em coin register}.
\begin{algo}{8cm}
\textbf{Generic Algorithm$(r,D,\Phi)$}\smallskip
\begin{enumerate}\setlength{\itemsep}{0pt} 
\item Create the state $\sum_{A\subset S : \size{A}=r}\ket{A}$
in the set register
\item Set up $D$ on $A$ in the data register
\item Create a uniform superposition over elements of $S-A$ in the coin register
\item Do $\Theta(n/r)^{k/2}$ times\vspace*{-\parsep}
\begin{enumerate}\setlength{\itemsep}{0pt} 
\item If $\Phi(D(A))$ accepts then do the phase flip, otherwise do nothing
\item Do $\Theta(\sqrt{r})$ times \textbf{Quantum Walk} 
updating the data register
\end{enumerate}\vspace*{-\parsep}
\item If $\Phi(D(A))$ rejects then reject, otherwise output
the collision given by $\Phi(D(A))$.
\end{enumerate}
\end{algo}

\begin{theorem}[\cite{amb04}]\label{ambainis1}
\textbf{Generic Algorithm} solves \textsc{Unique $k$-Collision}
with some positive constant  probability
and has query complexity 
$O(s(r)+(\tfrac{n}{r})^{k/2}\times (c(r)+\sqrt{r}\times u(r)))$.
\end{theorem}

Moreover it turns out that, when \textsc{Unique $k$-Collision}
has no solution, \textbf{Generic Algorithm} always rejects, and
when \textsc{Unique $k$-Collision} has a solution $c$, \textbf{Generic Algorithm}
outputs $c$ with probability $p=\Omega(1)$ which only depends on $k$, $n$ and $r$.
Thus using quantum amplification, one can modify \textbf{Generic Algorithm}
to an exact quantum algorithm.
\begin{corollary}\label{ambainis2}
\textsc{Unique $k$-Collision} can be solved with probability $1$ in
quantum query complexity 
$O(s(r)+(\tfrac{n}{r})^{k/2}\times (c(r)+\sqrt{r}\times u(r)))$.
\end{corollary}

One can make a random reduction from \textsc{Collision} to
\textsc{Unique Collision} if the definition on $\Phi$ is slightly generalized.
We add to the input of the checking procedure
a relation $\mathcal{R}\subseteq S^k$ which restricts the collision set $\mathcal{C}$ to $\mathcal{C}\cap\mathcal{R}$.
The reduction goes in the standard way
using a logarithmic number of randomly chosen relations $\mathcal{R}$, and hence an additional logarithmic factor
appears in the complexity.
If the collision relation is robust in some sense, one can improve this reduction by removing the $\log$ factors
(see for example the reduction used by Ambainis in~\cite{amb04}).

\begin{corollary}\label{ambainis0}
\textsc{Collision} can be solved in quantum
query complexity $$\tilde{O}(s(k)+\tfrac{n}{k}\times (c(k)+\sqrt{k}\times u(k))).$$
\end{corollary}

The tables below summarize the use of
the above formula for various problems.

{\footnotesize
\begin{center}\begin{tabular}{l|l}
{\em Problem} & {\em Collision relation} \\\hline
\textsc{Element}&\\
\textsc{distinctness} & $(u,v)\in \mathcal{C}$ iff $u\neq v$ and $f(u) = f(v)$ \\
\textsc{Graph}&\\
\textsc{Collision($G$)}  & $(u,v)\in \mathcal{C}$ iff $f(u) = f(v) = 1$
and $(u,v)\in G$ \\
&\\
\textsc{Triangle}  & $(u,v)\in \mathcal{C}$ iff there is a triangle $(u,v,w)$ in $G$
\end{tabular}
\end{center}

\begin{center}
\begin{tabular}{l|l|l|l}
& 
{\it Setup cost} & {\it Update cost} & {\it Checking cost} \\
{\em Problem} &$s(r)$&$u(r)$&$c(r)$\\\hline
\textsc{Element} &&&\\
\textsc{distinctness} & $ r$ & $ 1$ & $ 0$\\
\textsc{Graph}&&&\\
\textsc{Collision($G$)}  & $ r$ & $1$ & $ 0$\\
&&&\\
\textsc{Triangle}  & $O(r^2)$ & $ r$ & $ {O}(r^{2/3}\sqrt{n})$\\
\end{tabular}
\end{center}
}
\subsection{Graph Collision Problem}
Here we deal with an interesting variant of \textsc{Collision}  which will be also useful for
finding a triangle. The problem is parametrized by some graph $G$
on $n$ vertices which is given explicitly. 
\begin{quote}
\textsc{Graph Collision($G$)}\\
\textit{Oracle Input:} A boolean function  $f$ on $[n]$ which defines
the relation $\mathcal{C} \subseteq [n]^2$ such that $\mathcal{C}(u,u')$ iff
$f(u)=f(u')=1$ and $(u,u')\in E$. \\
\textit{Output:} 
A pair $(u,u')\in\mathcal{C}$ if it is non-empty,
otherwise reject.
\end{quote}
Observe that an equivalent formulation of the problem is to decide if the set of vertices
of value 1 form an independent set in $G$.

\begin{theorem}\label{graph-collision}
\textsc{Graph Collision($G$)} can be solved with positive constant  probability
in quantum
query complexity $\tilde{O}(n^{2/3})$.
\end{theorem}
\begin{proof}
We solve \textsc{Unique Graph Collision($G$)} using
Corollary~\ref{ambainis0},  with $S=[n]$ and $r=n^{2/3}$.
For every $U\subseteq [n]$ 
we define $D(U)=\{(v,f(v)):v\in U\}$, and let
$\Phi(D(U))=1$ if there are $u,u'\in U$ that satisfy
the required property.
Observe that $s(r)=r$, $u(r)=1$ and $c(r)=0$.
Therefore we can solve the problem in quantum query complexity
$\tilde{O}(r+\tfrac{n}{r}(\sqrt{r}))$
which is $\tilde{O}(n^{2/3})$ when $r=n^{2/3}$.
\end{proof}

\subsection{Triangle Problem}

\begin{theorem}\label{triangle}
\textsc{Triangle} can be solved with positive constant  probability
in quantum query complexity $\tilde{O}(n^{13/10})$.
\end{theorem}
\begin{proof}

We use Corollary~\ref{ambainis0}
where $S=[n]$, $r=n^{2/3}$, and $\mathcal{C}$ is the set of triangle edges. 
We define $D$ for every $U\subseteq [n]$ by $D(U)=G|_U$,
and $\Phi$ by $\Phi(G|_U)=1$ if 
a triangle edge is in $G|_U$.
Observe that $s(r)=O(r^2)$ and $u(r)=r$. 
We claim that $c(r)=\tilde{O}(\sqrt{n}\times r^{2/3})$.

To see this, let $U$ be a set of $r$ vertices such that $G|_U$ is explicitly known,
and let $v$ be a vertex in $[n]$. We define an input oracle for 
\textsc{Graph Collision($G|_U$)} by $f(u) = 1$ if $(u,v)\in E$. 
The edges of $G|_U$ which together with $v$
form a triangle in $G$ are the solutions of \textsc{Graph Collision($G|_U$)}.
Therefore finding a triangle edge, if it is in $G|U$, can be done in quantum query
complexity $\tilde{O}(r^{2/3})$ by Theorem~\ref{graph-collision}.
Now using quantum amplification~\cite{bhmt02}, we 
can find a vertex $v$, if it exists, which forms a triangle with some edge of
$G|_U$, using only
$\tilde{O}(\sqrt{n})$ iterations of the previous procedure,
and with a polynomially small error (which has no influence in the whole algorithm).


Therefore, we can solve the problem in quantum query complexity
$\tilde{O}(r^2+\tfrac{n}{r}(\sqrt{n}\times r^{2/3}+\sqrt{r}\times r))$
which is $\tilde{O}(n^{13/10})$ when $r=n^{3/5}$.
\end{proof}

\subsection{Monotone Graph Properties with Small Certificates}
Let now consider the property of having a copy of
a given graph $H$ with $k>3$ vertices.
Using directly Ambainis' algorithm, one gets
an algorithm whose query complexity is $\tilde{O}(n^{2-2/(k+1)})$.
In fact we can improve this bound to $\tilde{O}(n^{2-2/k})$.
Note that only the trivial $\Omega(n)$ lower bound is known.
This problem was independently considered by Childs and Eisenberg~\cite{ce03} whenever
$H$ is a $k$-clique. 
Beside the direct Ambainis' algorithm, they obtained an $\tilde{O}(n^{2.5-6/(k+2)})$
query algorithm. For $k=4,5$, this is faster than the direct Ambainis' algorithm, but slower than ours.

\begin{theorem}\label{h-copy}
Finding in a graph a copy of a given graph $H$, with $k>3$ vertices,
can be done with quantum query complexity 
$\tilde{O}(n^{2-2/k})$.
\end{theorem}
\begin{proof}
We follow the structure of the proof of Theorem~\ref{triangle}.
We distinguish an arbitrary vertex of $H$.
Let $d$ be the degree of this vertex in $H$.

We say that a vertex $v$ and a set $K$ of $(k-1)$ vertices of $G$ are {\em $H$-compatible}
if 
the subgraph induced
by $K\cup\{v\}$ in $G$ contains a copy of $H$, 
in which $v$ is the distinguished vertex. 
We also say that the set $K$ is an {\em $H$-candidate} when there exists a vertex $v$
such that $v$ and $K$ are $H$-compatible.
Our algorithm will essentially find a set that contains an $H$-candidate.


We define an instance of  $(k-1)$-\textsc{Collision},
where $S=[n]$,  and $\mathcal{C}$ is the set of $H$-candidates.
We define $D$ for every $U\subseteq [n]$ by $D(U)=G|_U$,
and $\Phi$ by $\Phi(G|_U)=1$ if $U$ is contains an $H$-candidate.
Again $s(r)=O(r^2)$ and $u(r)=r$. 
We now claim that $c(r)=\tilde{O}(\sqrt{n}\times r^{d/(d+1)})$.

The checking procedure uses a generalization of \textsc{Graph Collision}
to $d$-ary relations.
If some vertex $v$ of $G$ is fixed, then we say that
a subset $W\subseteq U$ of size $d$ is in relation if 
there exists $W\subseteq K\subseteq U$ such that
$v$ and $K$ are $H$-compatible in $G$, and
$v$ is connected to every vertex of $W$.
Following the arguments of the proof of Theorem~\ref{graph-collision}
(where the function $f$ takes the value $1$ on a vertex $u\in U$ if $(u,v)$
is an edge in $G$), 
we find a $d$-collision in quantum query complexity $\tilde{O}(r^{d/(d+1)})$
when it exists.
The checking procedure searches for a vertex $v$ for which this generalized
\textsc{Graph Collision} has a solution using a standard Grover search.

The overall parameterized query complexity is therefore
$$\tilde{O}\left(r^2+\left(\frac{n}{r}\right)^{(k-1)/2}
\left(\sqrt{n}\times r^{d/(d+1)}+\sqrt{r}\times r\right)\right).$$
By optimizing this expression (that is, by balancing the first and third terms), it turns out that
the best upper bound does not depend on $d$. Precisely the expression is optimal
with $r=n^{1-1/k}$, which gives the announced bound.
However, one can imagine a different algorithm for the checking procedure where
the choice of $d$ might be crucial.

To conclude, note that once a set $U$ of size $r$ that contains an $H$-candidate is found,
one can obtain a copy of $H$ in $G$ in the complexity of the checking cost $c(r)$.
\end{proof}

We conclude by extending this result for monotone graph properties 
which might have several small $1$-certificates.
\begin{corollary}\label{certificate}
Let $\varphi$ be a monotone graph property whose $1$-certificates
have at most $k>3$ vertices.
Then deciding $\varphi$, 
and producing a certificate whenever $\varphi$ is satisfied,
can be done with quantum
query complexity to the graph in $\tilde{O}(n^{2-2/k})$.
\end{corollary}

\section*{Acknowledgments}
We would like to thank Andris Ambainis for useful discussions and
for sending us a preliminary version of~\cite{amb04}.

\bibliographystyle{alpha}
\bibliography{mss03}

\end{document}